\documentclass[aps,prd,showpacs,showkeys,floatfix,nofootinbib,twocolumn
,10pt]{revtex4-1}
\usepackage{ragged2e}
\usepackage{ulem}
\usepackage{color}
\usepackage{graphicx}
\usepackage{epsfig}
\usepackage{bm}
\usepackage{amsthm}
\usepackage{amsfonts}
\usepackage{float}
\usepackage{amsmath,amssymb}
\usepackage{graphicx}
\usepackage[caption = true]{subfig}
\usepackage{hyperref}
\usepackage{cleveref}
\usepackage[utf8]{inputenc}
\DeclareMathOperator{\Tr}{Tr}

\begin{document}
\title{Consistent approach to study gluon quasi-particles}
\author{Chowdhury Aminul Islam}
\email{chowdhury.aminulislam@ucas.ac.cn}
\affiliation{School of Nuclear Science and Technology, University of 
Chinese Academy of Sciences, Beijing 100049, China}
\author{Munshi G. Mustafa}
\email{munshigolam.mustafa@saha.ac.in}
\affiliation{Theory Division, Saha Institute of Nuclear Physics,
1/AF, Bidhannagar, Kolkata 700064, India}
\affiliation{Homi Bhabha National Institute, Anushaktinagar, Mumbai, 
Maharashtra 400094, India}
\author{Rajarshi Ray}
\email{rajarshi@jcbose.ac.in}
\author{Pracheta Singha}
\email{pracheta.singha@gmail.com}
\affiliation{Center for Astroparticle Physics \& Space Science, Bose
Institute, Block-EN, Sector-V, Salt Lake, Kolkata-700091, India}  
\affiliation{Department of Physics, Bose Institute, 93/1, A. P. C 
Road, Kolkata - 700009, India}
%
%
\begin{abstract}
We discuss a novel approach to estimate the partition function in
effective model frameworks when the effective potentials have 
multiple
extrema, so that ascertaining a mean field becomes difficult. Using
this approach we present a consistent model to study the 
thermodynamic
properties of gluon quasi-particles as a function of temperature, 
both
in the color confined and the color deconfined phases.
\end{abstract}

\maketitle

\section{\label{sec:Intro} Introduction}
In the strong coupling regime the thermal physics of strongly
interacting matter is best described by quantum chromodynamics (QCD) on
space-time lattices~\cite{Boyd:1996bx, Engels:1999tz, Fodor:2001au,
Fodor:2002km, Allton:2002zi, Allton:2003vx, deForcrand:2003vyj,
Aoki:2006br, Aoki:2006we, Aoki:2009sc, Borsanyi:2013hza}.  Deconfinement
of quarks and gluons and the chiral symmetry restoration at crossover
temperatures $T_c \sim 150$ MeV is now well
documented~\cite{Bazavov:2011nk, HotQCD:2014kol, Borsanyi:2013bia}.  The
deconfinement transition in a pure glue system is however found to be of
first-order at temperatures $T_d \sim 270$ MeV~ \cite{POLYAKOV1978477,
McLerran:1981pb, Svetitsky:1982gs, Karsch:1989pn, Rothe:1992nt}. The
thermal average of the Polyakov loop in the fundamental representation
$\hat{L}_F$ gives the static quark free energy, and is considered as the
order parameter \cite{McLerran:1981pb, Svetitsky:1982gs}.  Polyakov loop
in the deconfined phase also breaks spontaneously the symmetry of the
gluon action under Z(3) twists on the gluon fields at the temporal
boundary.  Similarly, the Polyakov loop in the adjoint representation is
related to the free energy of a static adjoint color source $\hat{L}_A$
~\cite{Abramchuk:2018bco, Simonov:2010bf, Marsh:2013xsa, Gupta:2007ax}.
But it is always invariant under the Z(3) twists of the gluon fields at
the physical boundary. 

Significant efforts have been put in for building effective models for a
spontaneous Z(3) symmetry breaking, with $\hat{L}_F$ as order parameter,
using Landau type of polynomial potentials~\cite{ Meisinger:1995ih,
Pisarski:2000eq, Dumitru:2000in, Pisarski:2002ji, Dumitru:2004aa,
Braun:2010aa}.  Further models have been built~\cite{FUKUSHIMA2004277,
Megias:2004hj, PhysRevD.73.014019, Ghosh:2006qh, Megias:2006bn,
Mukherjee:2006hq, Ghosh:2007wy, Schaefer:2007pw} by introducing
effective $\hat{L}_F$ fields in lieu of background temporal gluon fields
in the chiral models like Nambu-Jona-Lasinio (NJL)
model~\cite{Nambu:1961tp, Nambu:1961fr, Hatsuda:1994pi, Vogl:1991qt,
Klevansky:1992qe, Buballa:2003qv} or Chiral Sigma
models~\cite{Jungnickel:1995fp, Berges:1998ha, Tetradis:2003qa,
Schaefer:2006ds, Schaefer:2006sr}. These Polyakov loop enhanced chiral
models give simple but insightful description of thermodynamics of
strong interactions~\cite{Kashiwa:2007hw, Fu:2007xc, Davoudiasl:2007wf,
Ciminale:2007sr, Costa:2008yh, Fukushima:2008wg, Buballa:2008ru,
Tsai:2008je, Sakai:2009vb, Braun:2009gm, Deb:2009ng, Costa:2009ae,
Schaefer:2009ui, Bhattacharyya:2010wp, Bhattacharyya:2010jd,
Bhattacharyya:2010ef, Morita:2011eu, Friesen:2011wt, Lourenco:2011buy,
Bhattacharyya:2011na, Inagaki:2012re, Inagaki:2012re,
Bhattacharyya:2012th, Bhattacharyya:2012rp, Bhattacharyya:2012up,
Megias:2013xaa, Ghosh:2014zra, Bhattacharyya:2014uxa, Ghosh:2014vja,
Bhattacharyya:2015kda, Bhattacharyya:2016jsn, Singha:2017ctc,
Bhattacharyya:2017gwt, Singha:2017jmq, Bhattacharyya:2019qhm}.  In these
models the gluon pressure is obtained from the polynomial thermodynamic
potential in ${\rm Tr}\hat{L}_F$.  But a more natural alternative seems
to be in terms of gluon quasi-particles~\cite{Peshier:1995ty,
Levai:1997yx, Peshier:1999ww, Meisinger:2003id, Plumari:2011mk},
including modifications due to the background
$\hat{L}_A$~\cite{Sasaki:2012bi, Ruggieri:2012ny, Islam:2012kv,
Alba:2014lda}. Here $\hat{L}_A$ is expected to induce statistical
confinement of gluons in a similar way as $\hat{L}_F$ does for quarks in
the Polyakov enhanced chiral models. But unfortunately the modified
statistics result in a negative gluonic pressure below $T_d$.  Various
authors have argued for additional terms to preserve overall positivity.
But the quasi-particle pressure itself still remains negative. This
lacunae may have slowed down further progress in this direction.

Here we argue that the issue lies with the method of obtaining the
statistics. Usually the saddle point approximation is employed to obtain
the mean value of the Polyakov loop, which is then put back to obtain
the thermodynamic potential. We propose a new prescription for obtaining
the thermal averages and thermodynamic observables that can solve the
issue and reliably predict various observables both below and above
$T_d$.

\section{\label{sec:formalism}Formalism}

\subsection{\label{sec:haar} Standard approach}
The Polyakov loop in the effective models is written in terms of the
background temporal gluon field $\mathcal{A}_0$, in the color 
diagonal
form as,
\begin{eqnarray}
\hat{L}_F \sim exp[i(\mathcal{A}_0^3T^3+\mathcal{A}_0^8T^8)/T].
\label{eq.polyaprx}
\end{eqnarray}
Here $T_3$ and $T_8$ are the diagonal generators of SU(3). 
Accordingly,
in terms of the class parameters $\theta_1$ and $\theta_2$ we have,
\begin{equation}
\hat{L}_F = diag(e^{i\theta_1},e^{i\theta_2},e^{-i(\theta_1+
\theta_2)}).
\label{funmatrix}
\end{equation}
The normalized character and its complex conjugate are then given 
as,
\begin{eqnarray}
\Phi =\frac{1}{3}\Tr\hat{L}_F
; \hspace{5mm}    
\bar{\Phi}=\frac{1}{3}\Tr\hat{L}^\dagger_F.
\label{funcharacter}
\end{eqnarray}

In general for SU($N_c$) the group invariant Haar measure $d\mu$ may 
be
expressed in terms of the distribution of eigenvalues as,
\begin{eqnarray}
\int d\mu 
&=& \frac{1}{N_c!}\left(\prod_{i=1}^{N_c} \int_0^{2\pi}
\frac{d\theta_i}{2\pi}\right) \delta\left(\sum_i \theta_i\right)
\prod_{i < j}|e^{i\theta_i} - e^{i\theta_j}|^2 \nonumber \\
&=& 1.
\label{eq.haarsunc}
\end{eqnarray}
For SU(3) the corresponding Haar measure is given by, 
\begin{equation}
\frac{1}{3!}\int_{0}^{2\pi}\int_{0}^{2\pi}
\frac{d\theta_1}{2\pi} \frac{d\theta_2}{2\pi} 
Det_{VdM}[\theta_1, \theta_2] =1,
\label{normalise}
\end{equation} 
where the Vandermonde determinant $Det_{VdM}$ is given as
(\cite{Islam:2012kv, Sasaki:2012bi, Ghosh:2007wy}),
\begin{eqnarray}
Det_{VdM} &=& 64
\sin^2\frac{(\theta_1-\theta_2)}{2} \sin^2\frac{(2\theta_1+
\theta_2)}{2}
\sin^2\frac{(\theta_1+2\theta_2)}{2} \nonumber \\
&=& 27
[1-6\bar{\Phi}\Phi+4(\bar{\Phi}^3+\Phi^3)-3(\bar{\Phi}\Phi)^2]~.
\label{haarphi}
\end{eqnarray}
Correspondingly, in both polynomial and quasi-particle  potentials, a
Vandermonde term can be included.  Thus the $n^{th}$ order polynomial
potential becomes,
\begin{eqnarray}
\Omega^{\prime}_{poly}=\left[\Omega_{poly}
\left(\alpha_{i=1 \cdots n}(T),\Phi,\bar{\Phi}\right)
+\kappa Det_{VdM}\right]T^4,
\label{eq.omegapoly}
\end{eqnarray}
where $\alpha_i$, $\kappa$ are model parameters.
The quasi-particle potential becomes,
\begin{equation}
\Omega_{gqp}^{\prime}=\Omega_{gqp}+\kappa Det_{VdM}T^4,
\label{eq.omegatot}
\end{equation}
where, 
\begin{eqnarray}
\Omega_{gqp}=
&=& {2T\int{\frac{d^3p} {(2\pi)^3} 
\ln\det{\big(1-\hat{L}_A e^{-\frac{|\vec{p}|}{T}}\big)}}} \nonumber \\
&=& {2T\int{\frac{d^3p} {(2\pi)^3}
\ln\left(1+\sum^{8}_{n=1} a_n e^{-\frac{n |\vec{p}|}
{T}}\right)}}~.
\label{eq.simpleg}
\end{eqnarray}
The coefficients $a_n$ for $n= 1 \cdots 8$, are,
\begin{eqnarray}
&a_8 &= 1\nonumber;~~
a_1 = a_7=1-9\bar{\Phi}\Phi \nonumber \\
&a_2 &= a_6=1-27\bar{\Phi}\Phi +27(\bar{\Phi}^3+\Phi^3) \nonumber \\
&a_3 &= a_5=-2+27\bar{\Phi}\Phi-81(\bar{\Phi}\Phi)^2 \nonumber \\
&a_4 &= 2[-1+9\bar{\Phi}\Phi-27(\bar{\Phi}^3+\Phi^3)+
81(\bar{\Phi}\Phi)^2]~,
\label{polya_distri}
\end{eqnarray}
The adjoint Polyakov loop is given as,
\begin{multline}
\hat{L}_A = diag\left(1, 1, e^{i(\theta_1-\theta_2)},
e^{-i(\theta_1-\theta_2)}, e^{i(2\theta_1+\theta_2)}, \right. \\ 
\left.e^{-i(2\theta_1+\theta_2)}, e^{i(\theta_1+2\theta_2)},
e^{-i(\theta_1+2\theta_2)}\right)~,
\label{adjLphi}
\end{multline}
with the corresponding normalized character
\begin{equation}
\Phi_A=\frac{1}{N_c^2-1}\Tr\hat{L}_A
=\frac{1}{8}\left({9\Phi\bar{\Phi}-1}\right)
~.
\label{adjcharacter}
\end{equation}

Given $\Omega$ from either Eq.~\ref{eq.omegapoly} or
Eq.~\ref{eq.omegatot}, one can solve for
\begin{eqnarray}
\frac{\partial \Omega}{\partial \Phi} =0 ~;~ 
\frac{\partial \Omega}{\partial \bar{\Phi}}=0~,
\label{saddle}
\end{eqnarray}  
obtaining the saddle point estimate for the mean fields $\Phi_{mf}$ and
$\bar{\Phi}_{mf}$, and the mean thermodynamic potential $\Omega = 
\Omega (\Phi_{mf},\bar{\Phi}_{mf})$.

In the quasi-particle picture this mean field approach gives
satisfactory results for $T>T_d$~\cite{Sasaki:2012bi, 
Ruggieri:2012ny, Islam:2012kv, Alba:2014lda}. Below $T_d$ $\Phi_{mf}
=\bar{\Phi}_{mf} =0$, and the thermodynamic potential becomes,
\begin{eqnarray}
& \Omega_{gqp}&({\Phi_{mf},\bar{\Phi}_{mf} \rightarrow 0})=2T
\int\frac{d^3p}{(2\pi)^3}
\left[\text{ln}
\left(1-e^{\left(-\frac{3|\vec{p}|}{T}\right)}\right)^2 \right.
\nonumber \\
& & \left.
+\text{ln}
\left(1-e^{\left(-\frac{|\vec{p}|}{T}+\frac{2\pi i}{3}\right)}
\right)
+\text{ln}
\left(1-e^{\left(-\frac{|\vec{p}|}{T}-\frac{2\pi i}{3}\right)}
\right)
\right].
\label{phiOomega}
\end{eqnarray} 
The last two terms are positive, resulting in an overall temperature
dependent negative pressure for $T< T_d$.

In Ref.~\cite{Sasaki:2012bi} the authors proposed a hybrid approach
including glue-balls implemented as dilaton fields, resulting in an {\it
overall} positive pressure below $T_d$.  In Ref.~\cite{Ruggieri:2012ny},
an additional pure matrix interaction term was used along with
$\Omega_{gqp}^{\prime}$ and a Weiss mean field analysis was done. In a
related Weiss averaging procedure \cite{Tsai:2008je}, a model
parametrization was invoked similar to the polynomial potential. In
Ref.~\cite{Alba:2014lda}, a Bag term was introduced along with
$\Omega_{gqp}^{\prime}$ and a thermodynamic consistency analysis was
done. Truly each of these additional terms may have significant physical
inputs. But the negativity of gluon quasi-particle pressure below $T_d$
remains unaddressed. 

\subsection{\label{sec:haaralt} Alternate approach}

Here we propose to use a matrix model for the background Polyakov loop.
We begin with the corresponding partition function given as,
\begin{eqnarray}
Z_{PL} 
&=& \int{\mathcal{D}\theta_1\mathcal{D}\theta_2 
exp\left[-\frac{1}{T}\int d^3x 
\Omega_{gqp}[\theta_1({\bf x}),\theta_2({\bf x})]\right]} \nonumber \\
&=& \int\prod_{\bf x} \frac{1}{24\pi^2}
d\theta_1({\bf x}) d\theta_2({\bf x}) Det_{VdM} \nonumber \\
&~&exp\left[-\frac{1}{T}\int d^3x 
\Omega_{gqp}[\theta_1({\bf x}),\theta_2({\bf x})]\right].
\label{eq.zpart1}
\end{eqnarray}
where $\Omega_{gqp}$ is given in Eq.~\ref{eq.simpleg}. Additional terms
as in Refs.~\cite{Sasaki:2012bi, Ruggieri:2012ny, Alba:2014lda} could be
introduced but are not relevant for the physics discussed here. As we
shall see that in our approach, even this simple $Z_{PL}$ is sufficient
to describe the pure gauge lattice field theory data quite
satisfactorily.

The Polyakov loop is an oscillating function of $\theta_1$ and
$\theta_2$, and so will be the thermodynamic potential. Hence the
configurations away from the saddle point may have a significant
measure. In fact below $T_d$, where $\langle \Phi \rangle = 0$,
configurations with $|\Phi|\sim 1/3$ is most preferred
\cite{Islam:2012kv}. Here instead, we compute the partition function
$Z_{PL}$, by numerically integrating over all the finite periodic
interval of the $\theta_1$ and $\theta_2$ fields. The difficulty is with
taking the $V \rightarrow \infty$ limit, and this is why the saddle
point analysis is the usual choice. However a simplification arises by
noting that the effective action contains no derivatives of the
$\theta_1$ and $\theta_2$ fields. Therefore the configuration space can
be split up into $N \rightarrow \infty$ independent and equivalent
points, such that the partition function becomes,
\begin{eqnarray}
Z_{PL} &=& z_{PL}^N
\label{part_coarse}
\end{eqnarray}
where,
\begin{eqnarray}
z_{PL} &=& \int \frac{1}{24\pi^2}d\theta_1 d\theta_2
Det_{VdM} exp\left[-\frac{v}{T}
\Omega_{gqp}[\theta_1,\theta_2]\right].
\label{eq.z}
\end{eqnarray}
and $v$ is a parameter with the dimension of volume. This is the only
free parameter and may be suitably related with $T_d$, the only physical
scale in the finite temperature SU(3) pure gauge field theory. We assume
$v=(\beta_1/T_d)^3$, where $\beta_1$ is some constant.  We further scale
out the momentum variable as $|\tilde{\vec{p}}| = |\vec{p}|/T$, whereby
the partition function may be expressed in terms of the scaled
temperature $T/T_d$ as,
\begin{eqnarray}
z &=& \int \frac{1}{24\pi^2}d\theta_1 d\theta_2
Det_{VdM} \nonumber \\
&& exp\left[{-2\left(\frac{\beta_1T}{T_d}\right)^3 
\int{\frac{d^3\tilde{p}}
{(2\pi)^3}
\text{ln}\big{(}1+\sum^{8}_{n=1} a_n e^{-n |\tilde{\vec{p}}|}
\big{)}}}\right]~.
\label{eq.ztd}
\end{eqnarray}
The scaled pressure can be expressed in terms of the scaled temperature
as,
\begin{eqnarray}
p/T^4 &=& \left(\frac{T}{V}\ln Z_{PL}\right)/T^4 
= \frac{1}{N v} N\ln z_{PL}/T^3 \nonumber \\
&=& \ln z_{PL} /\left(\beta_1T/T_d\right)^3 ~.
\label{eq.pressure1}
\end{eqnarray}

\section{\label{sec:result} Results}

\subsection[parameter]{Parameter Fitting}

\begin{figure}[!htb]
\centering
\includegraphics[scale=0.5]{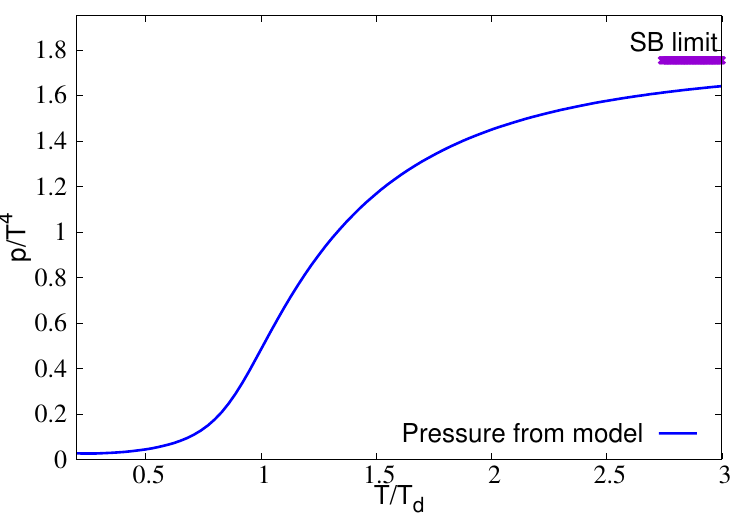}
\caption{\small{Scaled pressure of gluon quasi-particles as function
of scaled temperature.}}
\label{fg.press_com}
\end{figure} 
The variation of $p/T^4$ with $T/T_d$ obtained from
Eq.~\ref{eq.pressure1}, is completely consistent throughout the range of
temperatures (Fig.~\ref{fg.press_com}).  We chose $\beta_1 \sim 2$.
However for quantitative agreement with data from lattice SU(3) field
theory, we consider a temperature dependent effective mass ($m_g(T) $)
for the gluon quasi-particles. Effects of such constant mass was studied
in~\cite{Sasaki:2012bi}, while a temperature dependent ansatz was used
in~\cite{Ruggieri:2012ny}, following the studies
in~\cite{Peshier:2005pp}. We substitute $|\tilde{\vec{p}}|$ with
$\tilde{E}_g = \sqrt{|\tilde{\vec{p}}|^2+\tilde{m}_g(T)^2}$ in
Eq.~\ref{eq.ztd}, where $\tilde{m}_g(T) = m_g(T)/T $.  The lattice
data~\cite{Giusti:2016iqr} for pressure was solved for the scaled masses
from the pressure equation $p/T^4|_{model}=p/T^4|_{lattice}$ to an
accuracy of $10^{-6}$ or better.  The momentum rescaling is undefined
for $T \rightarrow 0$. Also the numerical uncertainties were
insignificant only above $T/T_d \sim 0.45$. Therefore the partition
function at a given $T/T_d$ was normalized with the one at $T/T_d =
0.45$.

\begin{figure}[!htb]
\centering
\includegraphics[scale=0.5]{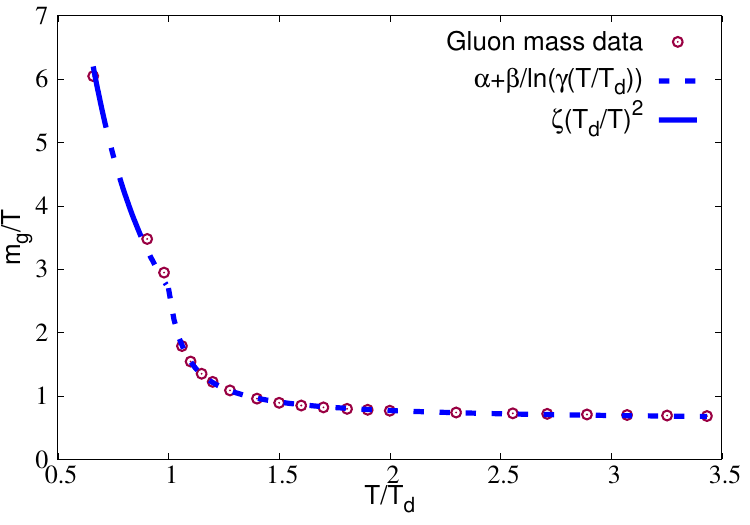}
\caption{\small{Scaled mass of gluon quasi-particles as function
of scaled temperature.}}
\label{fg.mass_pg}
\end{figure} 
The extracted scaled masses $\tilde{m}_g(T) = m_g(T)/T$ are shown in
Fig.\ref{fg.mass_pg} as data points. The functional dependence of
$\tilde{m}_g(T)$ has an abrupt change close to $T/T_d=1$, and may 
have a functional form,
\begin{eqnarray}
m_g(T)/T &=& \alpha + \beta/\ln(\gamma~T/T_d), \rm{~for~} T/T_d > 1 \\
         &=& \zeta~(T_d/T)^2, \rm{~for~} T/T_d < 1 .
\label{gluon_mass}
\end{eqnarray}
With $T_d$ arbitrary and assuming $v=(2/T_d)^3$, the functional fit is
shown in Fig.~\ref{fg.mass_pg}.  The parameters $\alpha$, $\beta$ and
$\gamma$ are obtained by least square fit using the "gnuplot" software,
and summarized in Table~\ref{table1}.
\begin{table}[!htb]
\begin{center}
	\begin{tabular}{|c|c|c|c|c|}
		\hline
		\multicolumn{4}{|c|}{Fitted}& \multicolumn{1}{|c|}{Chosen} \\
		\hline
		& \\[-3.3 ex]
$\alpha$ & $\beta$ & $\gamma$ & $\zeta$ & $v$(GeV$^{-3}$)\\ [1 ex]
		\hline
		& \\[-3.3 ex]
		0.548 & 0.174 & 1.083 & 2.70 & $(0.5T_d)^{-3}$ \\[1 ex]
		$\pm$0.006 & $\pm$0.005 & $\pm$0.004 & $\pm$0.07 & \\[1 ex]
		\hline
	\end{tabular}
	\caption{Model parameters}
	\label{table1}
	\end{center}
\end{table}

\begin{figure}[!htb]
{\includegraphics[scale=0.5]{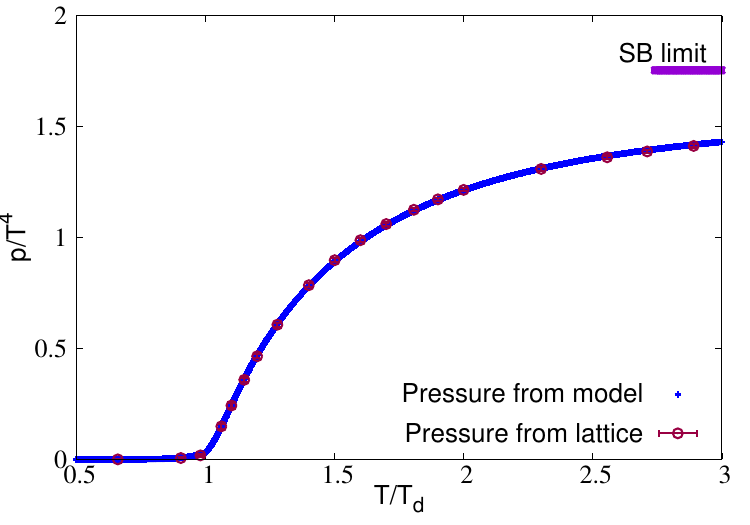}
\caption{Temperature variation of scaled pressure
(lattice data from~\cite{Giusti:2016iqr}).
}
\label{fg.press}}
\end{figure}
Given the mass parametrization a comparative plot of the scaled pressure
obtained in our model vis-a-vis the lattice QCD data is shown in
Fig.~\ref{fg.press}.

\subsection{\label{sec:thermo} Thermodynamic Observables}

Given $p/T^4$ as a function of $T/T_d$, other thermodynamic observables
like entropy density ($s$), energy density ($\epsilon$), specific heat
($c_V$) and speed of sound ($v_s$) may  be obtained as,
\begin{eqnarray}
& s/T^3 & 
=\frac{1}{T^3}\frac{\partial p}{\partial T} 
= \frac{1}{(T/T_d)^3}
\frac{\partial[ (p/T^4)(T/T_d)^4]}{\partial (T/T_d)},
\label{eq.entropy}\\
& \epsilon/T^4 &=p/T^4-s/T^3~,
\label{eq.energy}\\
& c_V/T^3 &=\frac{1}{T^3}\frac{\partial \epsilon}{\partial T}
= \frac{1}{(T/T_d)^3}
\frac{\partial[(\epsilon/T^4)(T/T_d)^4]}{\partial (T/T_d)},
\label{eq.spheat}\\
& v_s^2 &=\frac{\partial p}{\partial \epsilon}
=\frac{\partial p}{\partial T}/
\frac{\partial \epsilon}{\partial T}
=\frac{s/T^3}{c_V/T^3}~.
\label{eq.sos}
\end{eqnarray}

\begin{figure}[!htb]
\subfloat[]{\includegraphics[scale=0.5]{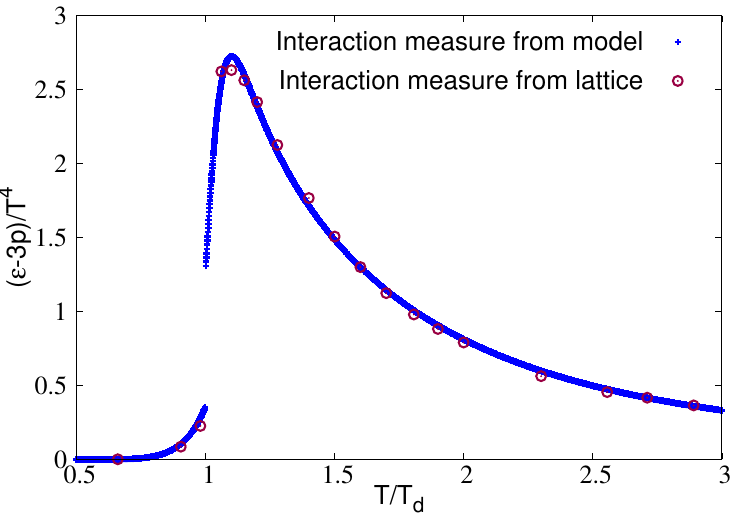}
	\label{fg.interaction}}\\
\subfloat[]{\includegraphics[scale=0.5]{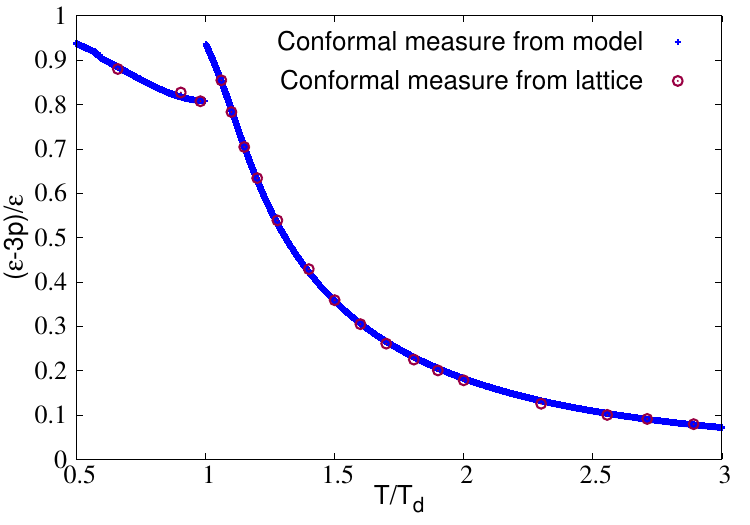}
	\label{fg.conformal}}
\caption{Temperature variation of (a) scaled interaction measure and
(b) scaled conformal measure (lattice data from~\cite{Giusti:2016iqr}).}
\label{fg.thermo1}
\end{figure} 
The scaled interaction measure (Fig.~\ref{fg.interaction}) is expected
to capture the deviation of thermal system from that of a relativistic
non-interacting gas of gluons. However for $T<<T_d$ due to the heavy
effective mass of the gluon quasi-particles as well as the
confinement-like interactions both $p/T^4$ and $\epsilon/T^4$ are
insignificant, and so is the interaction measure.  With increasing
$T/T_d$, both $m_g/T$ and the confinement effect decrease, thereby
increasing the interaction measure. A turnover occurs for $T/T_d>1$
inside the gluonic phase where the measure gradually decreases towards
relativistic ideal gas limit.

A more direct observable for transition from the non-relativistic
confined phase to the relativistic gluonic phase is the conformal
measure $(\epsilon-3p)/\epsilon$, which varies from 1 to 0 between the
two phases respectively (Fig.~\ref{fg.conformal}).  This behavior
follows the general trend of $m_g/T$.  At $T/T_d=1$ there is a sudden
gap arising out of the sudden changes in $m_g/T$ and the deconfining
effects.

\begin{figure}[!htb]
\subfloat[]{\includegraphics[scale=0.5]{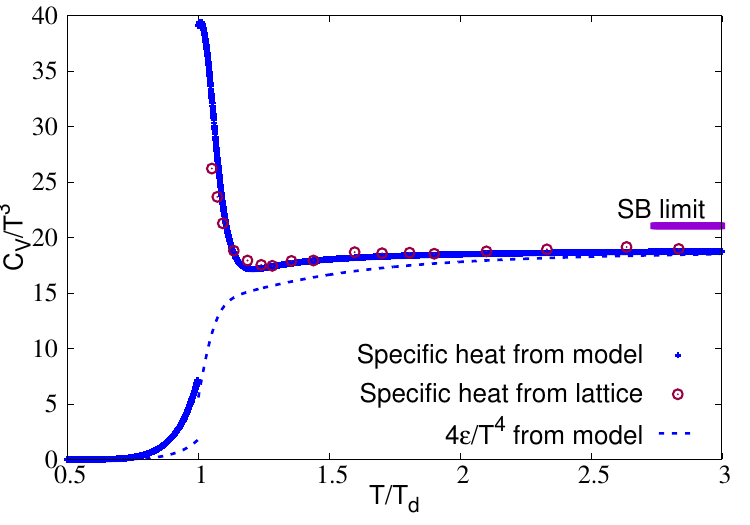}
	\label{fg.cv}}\\
\subfloat[]{\includegraphics[scale=0.5]{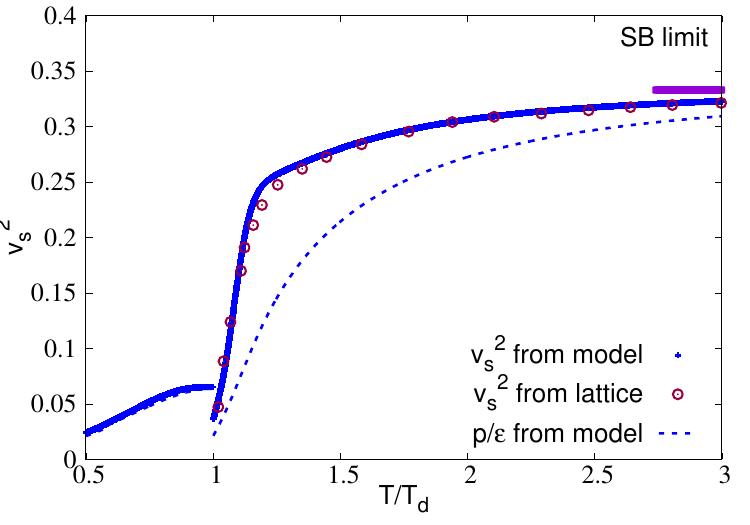}
	\label{fg.sss}}
\caption{Temperature variation of (a) 
scaled specific heat and (b) squared speed of sound (lattice data
from~\cite{Boyd:1996bx}).}
\label{fg.thermo2}
\end{figure} 

For a conformal theory in $d$ dimensions, $\epsilon=d.p$ and
$c_V/T^3=(1+d)\epsilon/T^4$. In Fig.~\ref{fg.cv} we show a direct
comparison of scaled specific heat with $\frac{4\epsilon}{T^4}$. The gap
in the $\epsilon/T^4$ gives an estimate of the latent heat of
transition. The scaled specific heat is both discontinuous and divergent
right at $T/T_d=1$. The agreement with lattice data could be ascertained
only for $T/T_d>1$ from the measurements reported in
Ref.~\cite{Boyd:1996bx}. 

Finally we present the behavior of the squared speed of sound $ (v_s^2)$
which is supposed to be an important transport coefficient determining
the hydrodynamic evolution in the heavy-ion collisions
\cite{Bhalerao:2005mm}. In the conformal limit $v_s^2=p/ \epsilon=1/3$.
A comparison of our estimation for the speed of sound along with the
ratio $p/\epsilon$ and measurements on the lattice~\cite{Boyd:1996bx} is
shown in~Fig.\ref{fg.sss}. Note that $p/\epsilon$ is three times the
additive inverse of the conformal measure. Reflection of such a
variation is seen in the $v_s^2 - T$ curve. The softest equation of
state is supposed to be at $T/T_d=1$, where $v_s^2$ drops towards zero.

Thus our model results for various sensitive thermodynamic observables
are in excellent numerical agreement with the lattice data.  In fact the
parametrizations obtained by fitting data from
Ref.~\cite{Giusti:2016iqr} (Fig.~\ref{fg.thermo1}), makes excellent
predictions for the data from Ref.~\cite{Boyd:1996bx}
(Fig.~\ref{fg.thermo2}).  

\subsection{\label{order}Order Parameter}

\begin{figure}[!htb]
	\centering
	\includegraphics[scale=0.5]{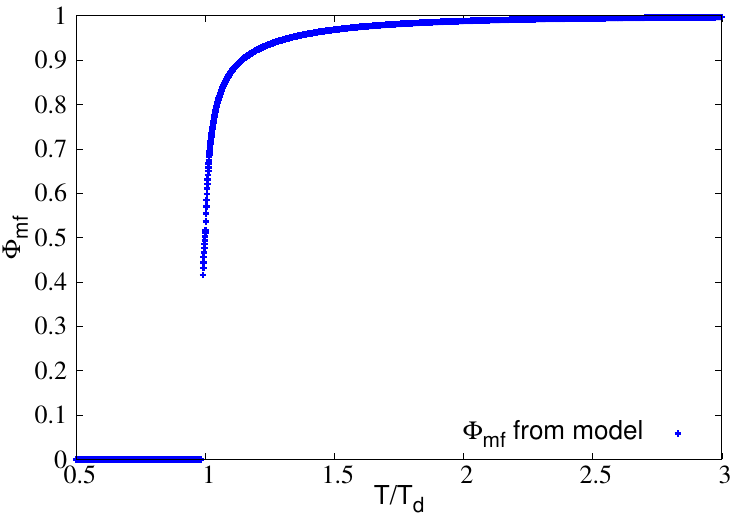}
	\caption{\small{Thermal evolution of $\Phi_{mf}$ (lattice data 
	from \cite{Gupta:2007ax}.}}
	\label{fg.saddle}
\end{figure}
As discussed earlier, $\langle \Phi \rangle $ is expected to vanish in
the Z(3) symmetric confined phase. In the deconfined phase the system
may be in any one of the spontaneously chosen ground states. For
numerical implementation, choosing the ground state requires some
biasing. For example with a source term for $\Phi$ towards one of the
ground states, one can consider the sequential limits $V\rightarrow
\infty$ and source term going to zero. However we have already
simplified the $V\rightarrow \infty$ limit. Consequently the thermal
averages of various operators are obtained as,
\begin{eqnarray}
&&\langle \hat{O}[\Phi,\bar\Phi] \rangle = 
\frac{1}{z}\int  d\theta_1 d\theta_2
Det_{VdM} O[\Phi,\bar\Phi]
\nonumber \\
&& exp\left[{-2\left(\frac{2T}{T_d}\right)^3\int{\frac{d^3\tilde{p}}
{(2\pi)^3}
\text{ln}\big{(}1+\sum^{8}_{n=1} a_n e^{-{n \tilde{E_g}}
}\big{)}}}\right]~.
\label{eq.expect2}
\end{eqnarray}
Since the averages are obtained by considering all the Z(3) states,
$\langle \Phi \rangle$ is trivially zero, both in the Z(3) symmetric and
symmetry broken phases. We are thus left only with the option of
saddle-point solution giving $\langle \Phi \rangle \simeq \Phi_{mf} $.
The $\Phi_{mf}$ is shown in Fig.~\ref{fg.saddle}, and shows the
characteristics of a first order phase transition.

\begin{figure}[!htb]
\includegraphics[scale=0.5]{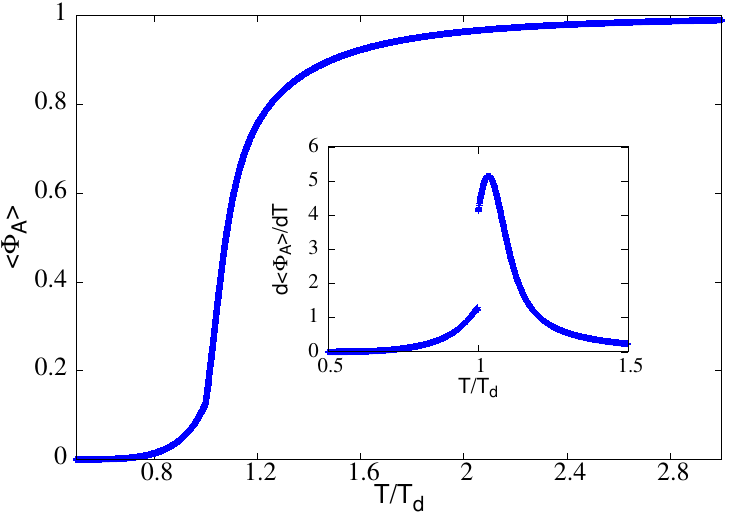}
\caption{Thermal evolution of $\langle\Phi_A\rangle$ and
$d\langle\Phi_A\rangle/dT$ (inset).}
\label{fg.adjoint}
\end{figure}
Unlike $\Phi$, $\Phi_A$ is invariant under the Z(3) transformation.
Therefore no complications arise in its evaluation. The temperature
dependence of $\langle\Phi_A\rangle$ is shown in Fig.~ \ref{fg.adjoint}.
Though a discontinuity corresponding to the one present in $m_g/T$
appears at $T/T_d=1$, the variation indicates a crossover rather than a
phase transition. This is further confirmed from the temperature
variation of the thermal derivative of $\langle\Phi_A\rangle$ (inset of
Fig.~\ref{fg.adjoint}), having a gap at $T/T_d=1$ and an inflection
point at some $T/T_d > 1$. This can be attributed to the fact that
within this current model framework, quasi-gluons have a finite mass at
temperatures below $T_d$. However a clear understanding of the relation
between the fundamental and adjoint representations of the Polyakov loop
is still to be investigated.

\begin{figure}[!htb]
	\centering
	\includegraphics[scale=0.5]{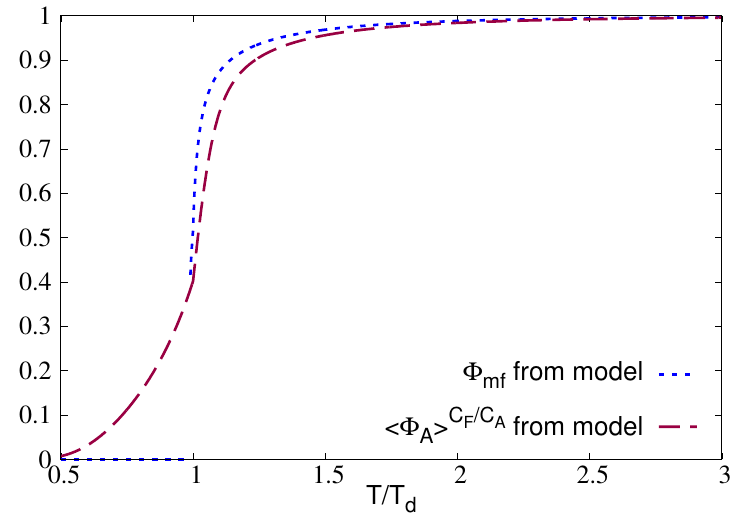}
	\caption{Casimir scaling as a function of $T/T_d$
(see~\cite{Gupta:2007ax} for notations).}
	\label{fg.casimir}
\end{figure}
Given that $\Phi$ is renormalization dependent on the lattice, it may
not agree with model results (as seen in Fig.~\ref{fg.saddle}). On the
lattice the Polyakov loop in various representations have been found to
follow Casimir scaling above $T/T_d=1$ \cite{Gupta:2007ax}. We find
$\langle \Phi \rangle$ and $\langle \Phi_A \rangle$ do follow the
scaling reasonably well for $T/T_d>1$ (Fig.~\ref{fg.casimir}).  \ Below
$T/T_d=1$, the scaling breaks because $\langle \Phi_A \rangle$ is
non-zero. This is natural as the quasi-gluons have finite mass.  and the
situation resembles the crossover observed in $\langle \Phi \rangle$ in
the presence of low mass quarks in models as well as on lattice
\cite{Clarke_2021}.  Possible effects of additional terms including
effects of glueballs, bag pressure or Polyakov loop interaction terms
discussed earlier may contribute to further understanding the behavior
of $\langle\Phi_A\rangle$ . These possibilities may be explored
elsewhere.

\section{Preliminary exploration including quarks}
The results discussed in the previous sections for the pure glue model
is expected to hold true in the presence of infinitely heavy quarks. For
practical purposes it should hold true even for a system of quarks whose
masses are much higher than the temperature scales. Here we make a
preliminary discussion on the presence of heavy quarks in the present
model. Neglecting effects of chiral physics altogether and incorporating
the Polyakov loop modified quark quasiparticle contribution we have the
additional potential, 
\begin{eqnarray}
\Omega_{qqp}=&2 N_f T \int \frac{d^3\tilde{p}}{(2\pi)^3}
\Bigg\lbrace\ln\Big[
3(\Phi+\bar{\Phi}e^{-\tilde{E}_q})e^{-\tilde{E}_q}+1+
\nonumber\\
& e^{-3\tilde{E}_q} \Big]+\ln\Big[3(\bar{\Phi}+
\Phi e^{-\tilde{E}_q}) e^{-\tilde{E}_q}+1+e^{-3\tilde{E}_q}
\Big]\Bigg\rbrace,
\nonumber\\
&~
\end{eqnarray}
where, $\tilde{E}_q=\sqrt{\tilde{p}^2+\frac{m_q}{T}^2}$, $m_q$ being the
quark mass. $N_f$ is number of quark flavors, which we shall consider to
be 2. The full potential will be the sum of $\Omega_{qqp}$ and
$\Omega_{gqp}$ (given in Eq.~\ref{eq.simpleg}) and to be used in
Eq.~\ref{eq.zpart1}. In this preliminary study with quarks we assume all
parameters of $\Omega_{gqp}$ to remain unchanged from those obtained in
the preceeding sections, except that we now specify $T_d=270$ MeV. We
shall discusss the behavior of the $\Phi_{mf}$ as a function of
temperature for various quark masses. Again for simplicity we shall use
$\Omega_{qqp}$ for various $m_q$, some of which are smaller than $T$.

\begin{figure}[!htb]
	\centering
	\includegraphics[scale=0.5]{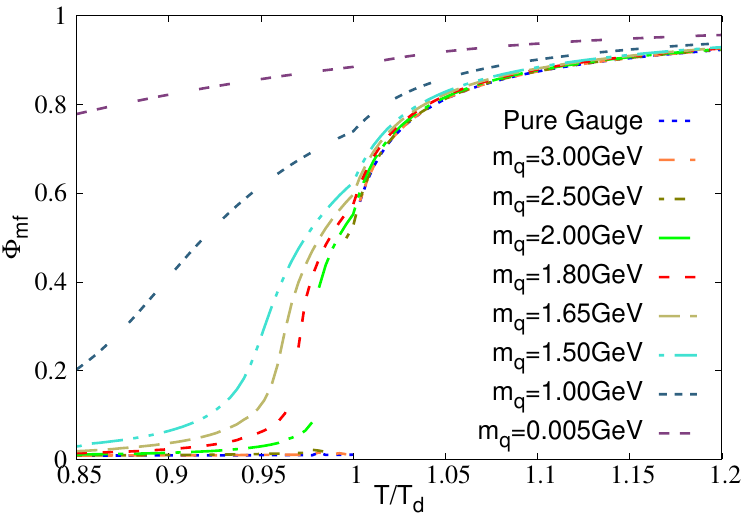}
	\caption{\small{Comparison of $\Phi_{mf}$ with varying quark masses.}}
	\label{fg.saddlecomp}
\end{figure}
The effect of introducing the quarks can be seen from
Fig.~\ref{fg.saddlecomp}. For $m_q = 3$ GeV the results are identical
with the pure gauge results. With reducing masses we find the
corresponding deconfinement temperature, as well as the gap of
$\Phi_{mf}$ to decrease. Subsequently between $1.65$ Gev $<m_q<$ $1.8$
GeV the transition goes over to a crossover. These results are
commensurate with the findings in the literature~\cite{Kashiwa:2012wa}.
The variation of $\Phi_{mf}$ with $T$ shows a dimple at the $T =
T_d$ of the pure glue model, which is nothing but an artefact of the
discontinuity of the variation of $m_g$ with $T$.  This can be taken
care of in a detailed analysis with dynamical quarks that will be
explored elsewhere.

\section{\label{sec:discussion}Discussion}
The gluon quasi-particle models are usually found to become inconsistent
in the confined phase. We identified the problem to lie with the saddle
point method. To overcome the problem we discussed a novel prescription
of obtaining the thermodynamic observables by a pseudo path integral
formalism. Essentially instead of considering only the saddle point
solution for the field variable, all possible field variables are
considered with their appropriate thermal weight functions. By
implementing this approach we predicted a variety of sensitive
thermodynamic quantities to a high level of accuracy.  In addition, we
observed that while the temperature variation of $\Phi$ indicates a
first order phase transition, that of $\Phi_A$ is almost like a
crossover. The latter seems natural as it is similar to thermal
variation of $\Phi$ when the quark masses are finite in chiral models.
However the deeper connection between the different representations of
the Polyakov loop in our model would need further investigation. A
preliminary study including quarks with heavy masses give consistent
results with existing literature.

We conclude that the quasi-particle model presented here is the most
consistent one for studies of color deconfinement and gluon
thermodynamics from cosmology to heavy-ion collision experiments. 

\begin{acknowledgments}

The authors would like to thank Department of Science and Technology
(DST) and Department of Atomic Energy (DAE), Govt. of India.  PS and CAI
would like to thank Leonardo Giusti and Michele Pepe for providing
useful information on lattice calculations in
Ref.~\cite{Giusti:2016iqr}. PS would like to thank Frithjof Karsch,
Anirban Lahiri and Sayantan Sharma for helpful discussions. CAI
acknowledges the facilities provided by the Tata Institute of
Fundamental Research, Mumbai, India where a part of the work was done,
and his present institute, the University of Chinese Academy of
Sciences, China. He is funded by the Chinese Academy of Sciences
President’s International Fellowship Initiative under Grant No.
2020PM0064 and has partial financial support from the Fundamental
Research Funds for the Central Universities, China.

\end{acknowledgments}
\bibliography{ref}

\end{document}